\documentclass[epj,twocolumn,dvips]{webofc}
\usepackage[varg]{txfonts}   
\usepackage{color}
\usepackage{graphicx}
\begin{document}
\title{INTERFERENCE EFFECT BETWEEN NEUTRON DIRECT AND RESONANCE CAPTURE REACTIONS FOR NEUTRON-RICH NUCLEI}
\author{
\firstname{Futoshi} \lastname{Minato}
\inst{1}\fnsep\thanks{\email{minato.futoshi@jaea.go.jp}} \and
\firstname{Tokuro} \lastname{Fukui}\inst{1,2}\fnsep
}

\institute{
Nuclear Data Center, JAEA, Tokai, 319-1195, Japan
\and
INFN, Napoli, 80126, Italy
}

\abstract{%
Interference effect of neutron capture cross section between the compound and direct processes is investigated. The compound process is calculated by resonance parameters and the direct process by the potential mode. The interference effect is tested for neutron-rich $^{82}$Ge and $^{134}$Sn nuclei relevant to $r$-process and light nucleus $^{13}$C which is neutron poison in the $s$-process and produces long-lived radioactive nucleus $^{14}$C ($T_{1/2}=5700$ y). The interference effects in those nuclei are significant around resonances, and low energy region if $s$-wave neutron direct capture is possible. Maxwellian averaged cross sections at $kT=30$ and $300$ keV are also calculated, and the interference effect changes the Maxwellian averaged capture cross section largely depending on resonance position.}
\maketitle

\section{Introduction}
\label{intro}
The neutron capture reaction is one of the simplest nuclear reactions, but its importance is recognized in various fields. For example, the heavy elements existing in the nature are considered to be produced by a successive neutron capture in a neutron-rich environment (the leading candidates are the $s$-process and $r$-process). In the field of accelerators and reactors, the neutron capture generates unstable nuclei in the structure materials used for confining neutrons and quantum beams after their long-term operation. The production of unstable nuclei by (secondary) neutrons cannot be ignored in terms of the radioactive protection and the decommissioning process. Vice versa, the neutron capture has an potential to transmute a radioactive unstable one to stable nucleus in terms of reducing radioactive wastes produced in various nuclear sites, especially in nuclear reactors. Moreover, using prompt $\gamma$-rays resulted from the neuron capture, non-destructive analysis in materials is also performed. It is therefore important to know accurate neutron capture cross sections systematically for nuclei in the nuclear chart, in particular at low energy region where its cross section becomes large.

The neutron capture at low energies can be divided into two parts, that is the compound process and the direct process. Both of the processes emit $\gamma$-rays with the same energy if the initial and final states are identical. However, the compound process stops by a compound state before $\gamma$-emission one hand, the direct process directly goes to the final state by the electronic dipole radiation of incident neutron on the other hand. The compound process has a characteristic resonance structure in the cross section (see Fig. \ref{capture} , for instance). The strong peaks emerge when the incident neutron wave function smoothly connects with the interior wave function of $A+1$ nucleus according to the $R$-matrix theory \cite{Wigner,WignerEisenbud}. For most nuclei, the compound process occupies its dominant part of the neutron capture reaction. This is because the number of resonances reaches to hundreds to millions at a capture state energy, $E^*=E+S_n$, where $E$ and $S_n$ are the incident neutron energy and the neuron threshold of $A+1$ nucleus, respectively. As a consequence, the neutron capture cross section of the compound process ranges approximately from $10^{-1}$ to $10^5$ barn at neutron resonance regions. On the other hand, typical neutron capture cross section of the direct process is the order of $0.1$ to $1$ mb.

However,  it is not always the case that the compound process is larger than the direct one in case of light nuclei and magic nuclei, for which the number of resonances at the capture state energy is small. In this case, neutron capture cross section of the compound process becomes small and the direct process comes to play a comparable role to the compound one. One of the examples is $^{12}$C \cite{Nagai} and $^{16}$O \cite{Igashira}. The direct process of those nuclei is strong enough that it affects the evolution of stars and $s$-process. Besides light nuclei and magic nuclei, the direct process becomes significant in neutron-rich nuclei as well because the number of resonances available to form the compound state is expected to be small due to the small neutron threshold energy.  For this reason, the direct process in neutron-rich nuclei has been actively investigated~\cite{Matthews,Bonneau,Rauscher2010,Chiba2008}.

The total neutron capture cross section is described in general by summing the results of the compound and direct processes. However, it was discussed \cite{Lynn,Lovas} that there is an interference effect between these processes, similar to the potential scattering and compound one in the elastic channel. Such an interference effect was investigated in proton capture reaction \cite{Rolfs}, however, it has not been clarified if the interference really exists in the neutron capture reaction. The evaluated neutron nucleus reaction cross section library, JENDL \cite{JENDL}, therefore did not consider the effect. However, several recent experiment works have shown unique data distributions in the cross sections of $^{9}$Be($\gamma,n$) and $^{16}$O($n,\gamma$)$^{17}$O that one cannot reproduce only by the compound and direct processes. Mengoni and Otsuka showed~\cite{Mengoni2000} that the cross sections can be reproduced theoretically if they take into account the interference effect.

It is then natural to think of whether the interference effect also becomes significant in neutron-rich nuclei. In fact, the compound and direct processes become comparable in the vicinity of the neutron magic number~\cite{Rauscher2010, Chiba2008}, and therefore a clear interference effect is expected to be found around that mass region. We choose three nuclei as test ones, that is $^{13}$C, $^{86}$Ge, and $^{134}$Sn. The first nucleus locates next to the stability line in the nuclear chart, but it is important for nuclear data evaluation because $^{13}$C is neutron poison in $s$-process and can be a generator of long-lived nucleus $^{14}$C ($T_{1/2}=5700$ y). The latter two are typical neutron-rich nuclei which may be involved in $r$-process.

In addition to neutron direct capture reaction, there is a semi-direct process coming out of $E=5 \sim 10$ MeV. But we do not discuss it because our interest in this work is less than $E=1$ MeV.

This paper is organized as follows. In Sec. \ref{sec-2}, we describe the formalism to calculate the neutron capture cross section briefly. Sec. \ref{sec-3} gives our result and Sec. \ref{sec-4} summarizes this work and gives the future perspectives.

\section{Calculation}
\label{sec-2}
In neutron resonance region for the nuclei studied in this work, the exit channel is classified only to the elastic scattering and the capture reaction, and other channels do not take part in. We also assume that the target nucleus is spherical.

The cross section of neutron capture can be described by the $R$-matrix theory, which is given by \cite{Wigner,WignerEisenbud,Descouvemont,Lynn,Lane,Lovas}
\begin{equation}
\sigma(E)
=
\frac{\pi}{k^2}g_J\left|\sum_\lambda\frac{ie^{-i\phi}\Gamma_{\lambda n}^{1/2}\Gamma_{\lambda\gamma\mu}^{1/2}}{(E_\lambda-E)-i\Gamma_\lambda/2}\right|^2,
\label{BW1}
\end{equation}
where $\Gamma_\lambda=\Gamma_{\gamma n}+\Gamma_{\lambda \gamma \mu}$ is the total decay width for  resonance state $\lambda$, $\Gamma_{\lambda n}$ the neutron width, $\Gamma_{\lambda\gamma\mu}$ the $\gamma$ width, and $E_\lambda$ the resonance energy. The index $\mu$ indicates a single particle orbit of $A+1$ nucleus in which the incident neutron is settled after the prompt $\gamma$ radiation; $g_J=(2J_i+1)/(2s+1)(2s_n+1)$ is the statistical factor, where $s$ and $s_n$ are the spins of target and projectile, respectively, $\vec{J}_i=\vec{S}+\vec{l}_i$ the total spin of the system, $\vec{S}=\vec{s}+\vec{s}_n$ the total spin, and $l_i$ the relative orbital angular momentum between the projectile and the target. The phase shift for the potential scattering is defined by $\phi$. Incident neutron wave number is defined by $k=\sqrt{2mE}/\hbar$, where $E$ and $m=A/(A+1)m_N$ are the incident neutron energy and the reduced mass, respectively.

If each resonance is isolated from others, Eq.~\eqref{BW1} reads
\begin{equation}
\sigma_{BW}(E)=
\frac{\pi}{k^2}g_J\sum_\lambda\frac{\Gamma_{\lambda n}\Gamma_{\lambda\gamma\mu}}{(E_\lambda-E)^2+(\Gamma_\lambda/2)^2}.
\label{BW2}
\end{equation}
Equation \eqref{BW2} is the so-called the multilevel Breit-Wigner formula. The cross section has a maximum value at $E=E_\lambda$.

The direct radiative neutron capture cross section can be calculated by
\begin{equation}
\sigma_{dir}(E)=\frac{\pi}{k^2}g_J|U_{dir}|^2,
\end{equation}
where the collision matrix $U_{dir}$ is written as \cite{Lane,Lovas}
\begin{equation}
U_{dir}=\frac{1}{\sqrt{2J_i+1}} \left(\frac{16m}{9\hbar^2 k}\right)^{1/2} \bar{e} \sum_{fi} k_\gamma^{3/2} Q_{fi}.
\end{equation}
The factor $\sqrt{m/\hbar k}=v^{-1/2}$ is coming from the assumption of a unit-flux incoming wave in the entrance channel. The effective charge of neutron is $\bar{e}=-eZ/A$, the emitted $\gamma$-ray wave number $k_\gamma=\epsilon_\gamma/\hbar c$, $\epsilon_\gamma=E-\epsilon_f$ the $\gamma$-ray energy, $\epsilon_f$ the single particle energy of the final state calculated by the Woods-Saxon potential, and $Q_{fi}=\mathcal{T}_{fi}B_{fi}A_{fi}$ the transition matrix \cite{Mengoni2000}, where $B_{fi}$ is the parameter concerning the spectroscopic factor.  The two indices $f$ and $i$ are defined as $f\equiv (j_f, l_f)$ and $i\equiv (j_i, l_i)$, respectively. The radial integral part, $\mathcal{T}_{fi}$, is given by \cite{Mengoni1995, Mengoni2000}
\begin{equation}
\mathcal{T}_{fi}=\int_0^\infty \, u(j_fl_f;r) \, r \chi(j_il_i;r) \, dr,
\end{equation}
where the scattering wave function calculated by a neutron-nucleus optical potential is
\begin{equation}
\begin{split}
\chi(j_i l_i;r)
&=\sqrt{4\pi}\sqrt{2l_i+1}i^{l_i}\frac{i}{2}\left(I_{l_i}^{j_i}(r)-U_{l_i}^{j_i}O_{l_i}^{j_i}(r)\right)\\
&\xrightarrow[r\rightarrow\infty]{}
\sqrt{4\pi}\sqrt{2l_i+1}i^{l_i}\frac{i}{2}\left(e^{i(kr-\frac{l\pi}{2})}-U_{l_i}^{j_i} e^{-i(kr-\frac{l\pi}{2})} \right)
\end{split}
\end{equation}
and $u(j_fl_f;r)$ is the radial part of the single particle wave function of the final state. We here define the scattering matrix element $U_l^j=e^{-2i\phi}$. The angular momentum part, $A_{fi}$ is written as
\begin{equation}
\begin{split}
A_{fi}
&=\hat{J}_f\hat{j}_f \hat{j}_i \hat{l}_i \sqrt{\frac{3}{4\pi} }\langle l_i010|l_f0 \rangle\\
&\times\left\{
\begin{tabular}{ccc}
$j_i$    & $s$ & $J_i$\\
$J_f$ & $1$ & $j_f$
\end{tabular}
\right\}
\left\{
\begin{tabular}{ccc}
$l_i$ & $j_i$ & $1/2$\\
$j_f$ & $l_f$ & $1$
\end{tabular}
\right\}.
\end{split}
\end{equation}

So far, we considered the compound process and the direct process separately. The neutron capture cross section taking into account both processes simultaneously is given by~\cite{Lynn,Lovas}
\begin{equation}
\sigma_{n\gamma}(E)=\frac{\pi}{k^2}g_J\left|ie^{-i\phi}\sum_\lambda\frac{\Gamma_{\lambda n}^{1/2}\Gamma_{\lambda\gamma\mu}^{1/2}}{(E_\lambda-E)-i\Gamma_\lambda/2}+U_{dir}\right|^2.
\label{cap1}
\end{equation}
If each resonance with the same capture state is isolated, Eq.~\eqref{cap1} becomes as we have done in Eq. \eqref{BW2}
\begin{equation}
\begin{split}
\sigma_{n\gamma}(E)\simeq
&\sigma_{BW}+\sigma_{dir}\\
&+\frac{\pi}{k^2}g_J\sum_\lambda\frac{\Gamma_{\lambda n}^{1/2}\Gamma_{\lambda\gamma\mu}^{1/2}
2Re\left(-ie^{i\phi}U_{dir}(E_\lambda-E-i\Gamma_\lambda/2)\right)}{(E_\lambda-E)^2+(\Gamma_\lambda/2)^2}.
\end{split}
\label{cap2}
\end{equation}
The first and second terms are indentical to the Breit-Wigner formula and the direct capture cross section, respectively, and the third term is nothing but the one from the interference effect.

Single particle wave functions and levels are calculated by the Woods-Saxon potential, in which the parameters are $R_n=1.236A^{1/3}$ fm, the diffuseness $a=0.62$ fm, the potential depth $V=-50$ MeV, and the $LS$ potential $V_{LS}=19.4/r\frac{df(r)}{dr}\vec{l}\cdot\vec{s}$ MeV. In case of $^{13}$C, we adjust $V$ for each resonance to reproduce the neutron capture cross section at thermal energy and $E=64$ keV, that is $V_{1p_{3/2}}=-38.6$, $V_{2s_{1/2}}=-51.4$, and $V_{1d_{5/2}}=-48.0$ MeV, retaining the other parameters. The scattering neutron wave function is calculated by the optical model with the Kunieda potential~\cite{Kunieda}. In order to take into account the non-local effect of the optical potential on the scattering wave function, the Perey and Buck method \cite{PereyBuck} with the non-local parameter $\beta=0.85$ fm is adopted.

Because $^{82}$Ge and $^{134}$Sn are the short half-life nuclei, the information about the resonance parameters evaluated from experiments is not available. It is not a simple task to calculate those parameters by a theoretical method. Therefore, we adopted, in this work, $\Gamma_{\lambda n}$, $\Gamma_{\lambda \mu}$ and $E_\lambda$ given in TENDL-2015~\cite{TENDL}, in which their resonance parameters are calculated on the basis of statistical properties regarding the nuclear excitations. In case of $^{13}$C, we used a newly evaluated resonance parameter of JENDL \cite{NobuyukiIwamoto}.

If resonance state is $J\ne0$, the final state has two possibilities in case of $E1$ radiation, that is $J\pm1$. In this case, we assume the equality in the $\gamma$ decay widths, namely $\Gamma_{\lambda\gamma\mu}=\Gamma_{\lambda\gamma}/2$. Note that the interference effect between the compound process and direct process occurs when the entrance and exit channels of the two processes are exactly identical.

In order to calculate the interference effect given in Eq.~\eqref{cap2}, we are confronted by the sign problem in the term proportional to $Re(-ie^{\phi} U_{dir}(E_\lambda-E-i\Gamma_\lambda/2))$. In this work, the following prescription is adopted; the interference effect on the neutron capture cross section in the first resonance of $^{13}$C ($J=2$) and $^{132}$Sn ($J=1/2$) works constructively below the resonance and destructively above it as found in the experimental data of $^{16}$O($n,\gamma$) cross section \cite{Mengoni2000}. Then, we apply the same phase to the interference term of the other resonances. In case of $^{82}$Ge of which the direct capture is dominated by the $p$-wave neutron, the same phase as $^{13}$C is adopted.

\section{Result}
\label{sec-3}

\begin{figure}[h]
\includegraphics[width=0.90\linewidth]{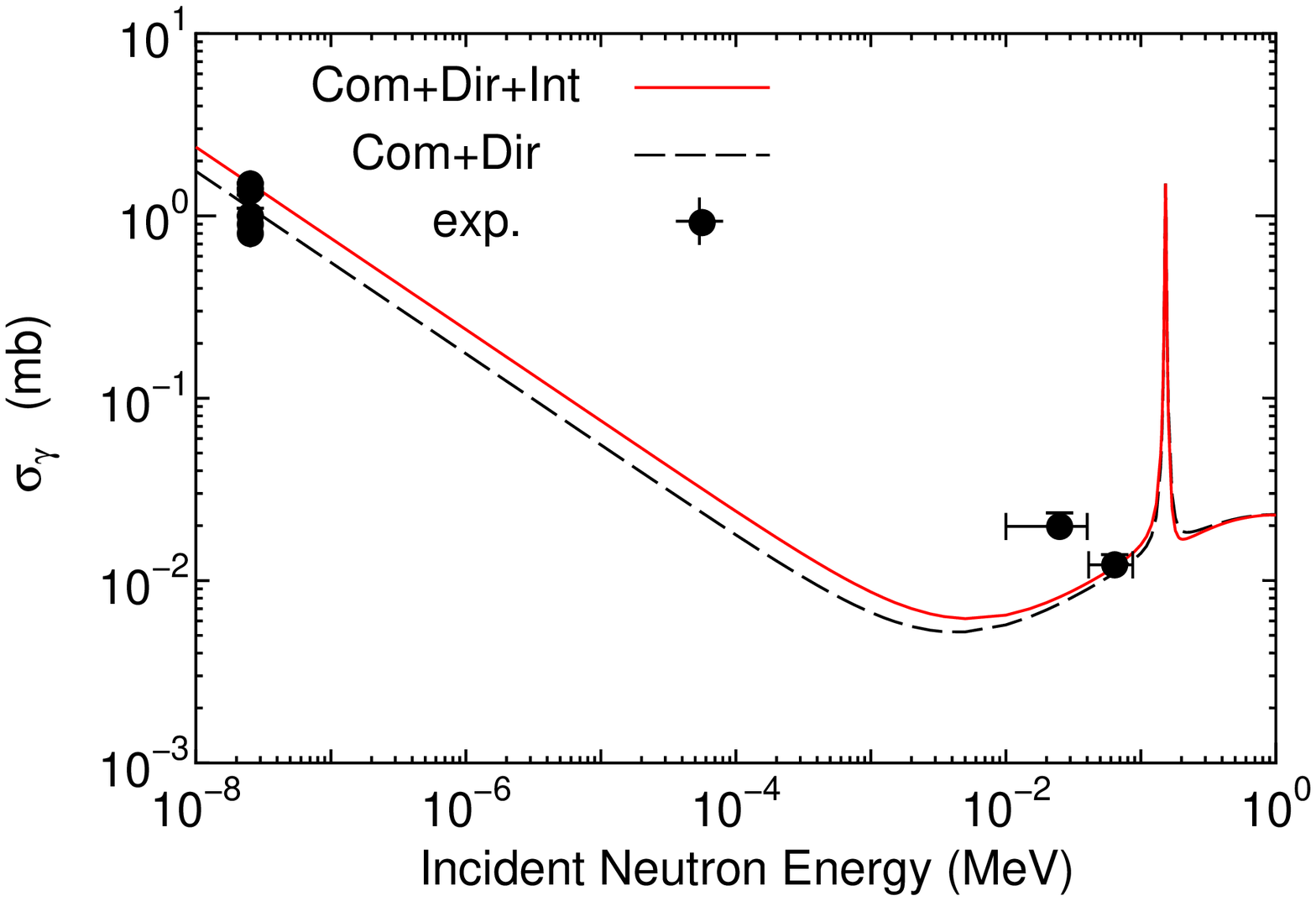}
\includegraphics[width=0.90\linewidth]{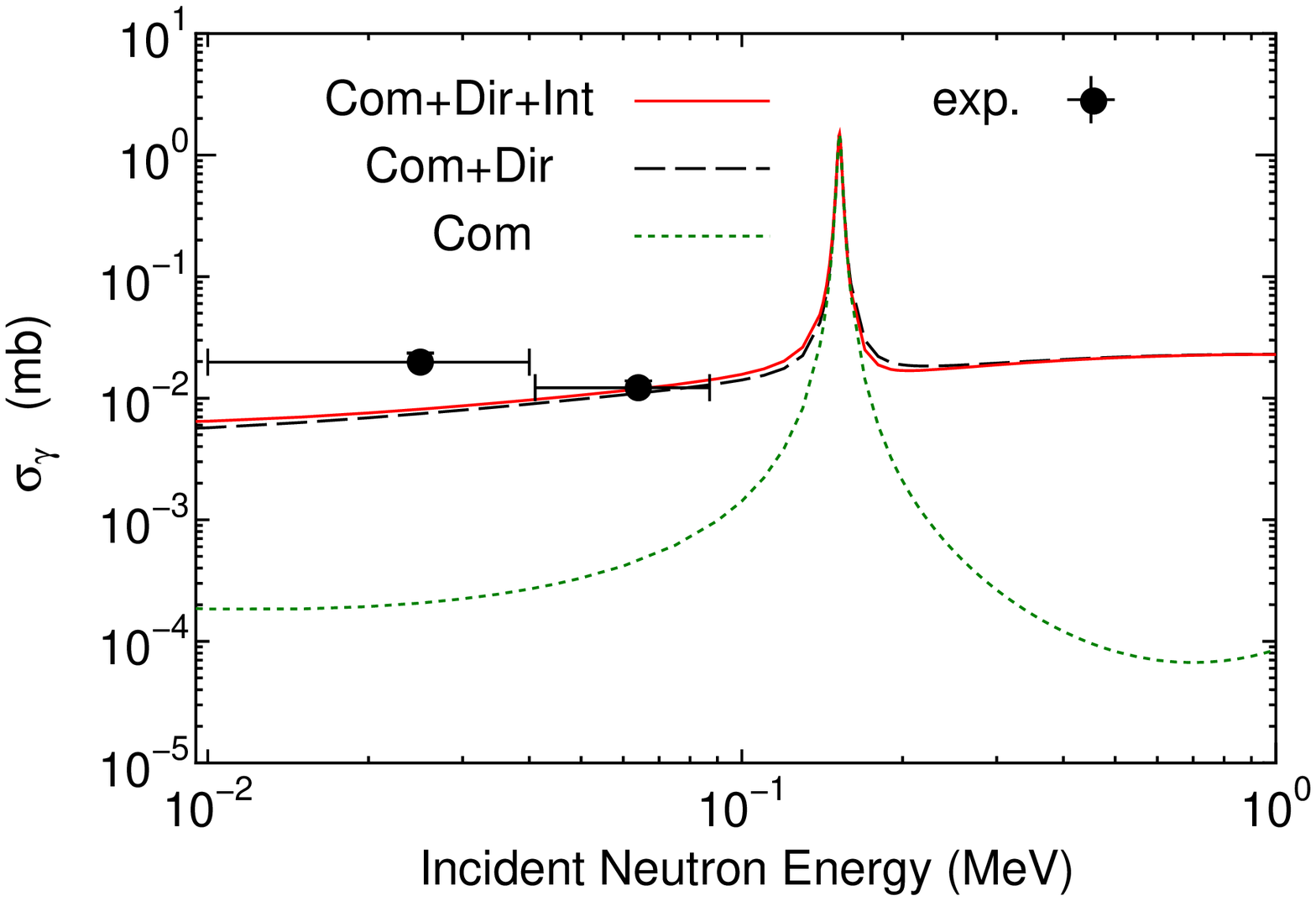}
\caption{Neutron capture cross section for $^{13}$C from $E=10^{-8}$ to $1$ MeV (top) and $E=0.01$ to $1$ MeV (bottom). The solid and dashed lines are the results of compound process + direct process + interference effect (Com+Dir+Int) and compound+direct process (Com+Dir), respectively. Note that the resonance parameter of $^{13}$C is taken from Ref. \cite{NobuyukiIwamoto}. The experimental data are taken from Ref. \cite{Firestone,Shima,exfor}.}
\label{capture0}
\end{figure}

Figure \ref{capture0} shows the neutron capture cross sections of $^{13}$C. The resultant $^{14}$C nucleus has the excited states of $2^+$ and $1^-$ below the neutron threshold energy, and the $0^+$ ground state. Since the target has the initial spin of $J^\pi=1/2^{-}$, the cross section is obtained by summing the results of total angular momentum, $J_i=0$ to $3$ based on the assumption that only E1 transition occurs. The top panel shows the cross section from $E=10^{-8}$ to $1$ MeV. The solid and dashed lines are the results of compound process + direct process + interference effect (Com+Dir+Int) and compound process + direct process (Com+Dir), respectively. The Woods-Saxon potential parameter $V$ is determined so as to reproduce the most recent experimental data of the thermal neutron capture cross section, $\sigma_\gamma(th)=1.496$ b \cite{Firestone} and at $E=64$ keV \cite{Shima} in case of Com+Dir+Int, as mentioned in the previous section. The interference effect plays a significant role from low energy region. This is because the $s$-wave neutron capture into the $1p_{1/2}$ bound state is concerned both with the compound and direct processes, and those cross sections are comparable. As incident neutron energy increases, the contribution from the $s$-wave neutron becomes weak, and the $p$-wave neutron comes to play a role from $E\sim0.01$ MeV instead.

In the bottom panel of Fig.~\ref{capture0}, the neutron capture cross section from $E=0.01$ to $1$ MeV is shown. The dotted line is the result of only the compound process. It is found that the direct process is essential to explain the experimental data \cite{Shima} around $10^{-2}$ to $10^{-1}$ MeV. However, the interference effect around the resonance at $E=152.4$ keV ($J=2$) is not strong as studied in $^{16}$O \cite{Mengoni1995}.

\begin{figure}[h]
\includegraphics[width=0.90\linewidth]{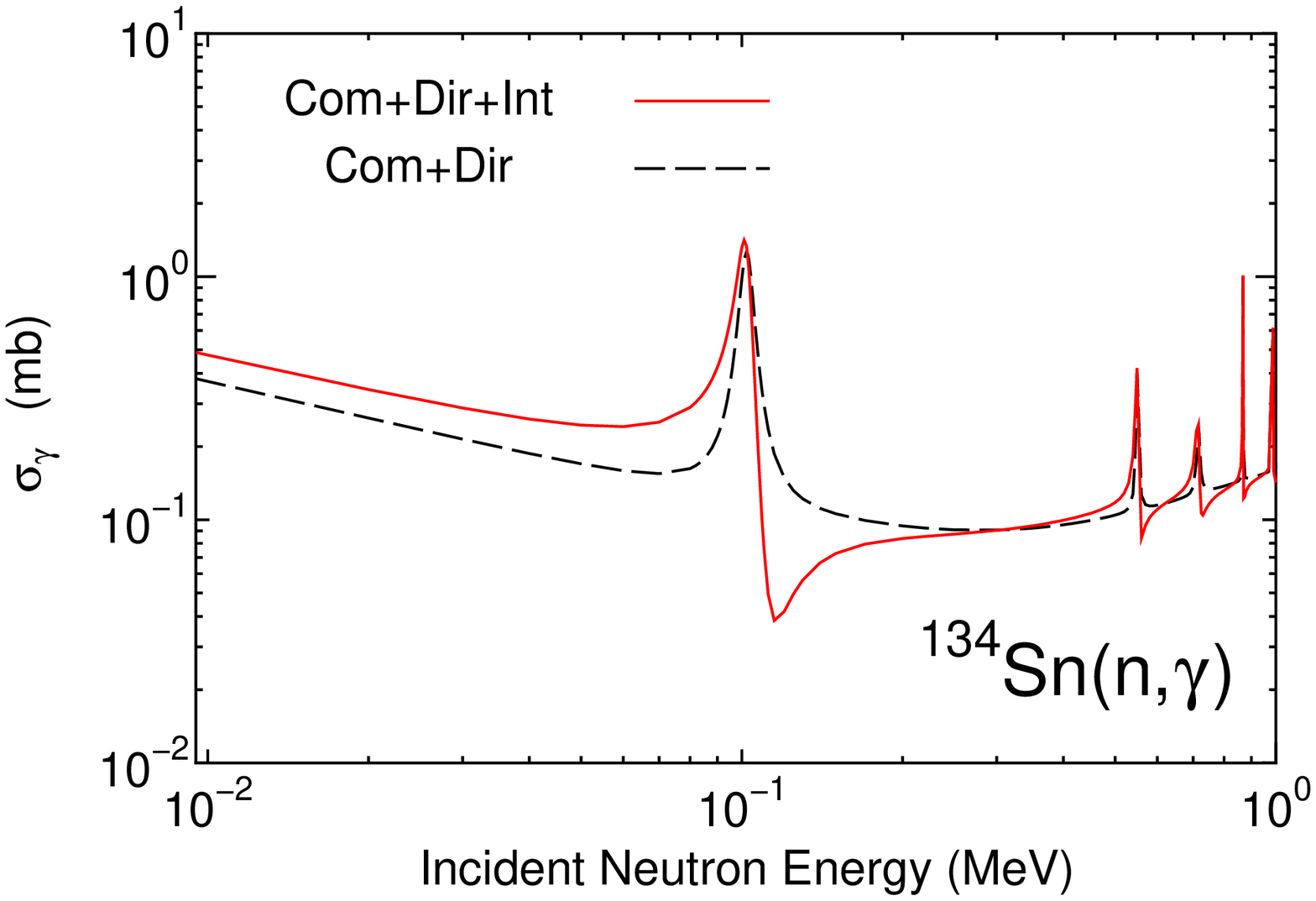}
\includegraphics[width=0.90\linewidth]{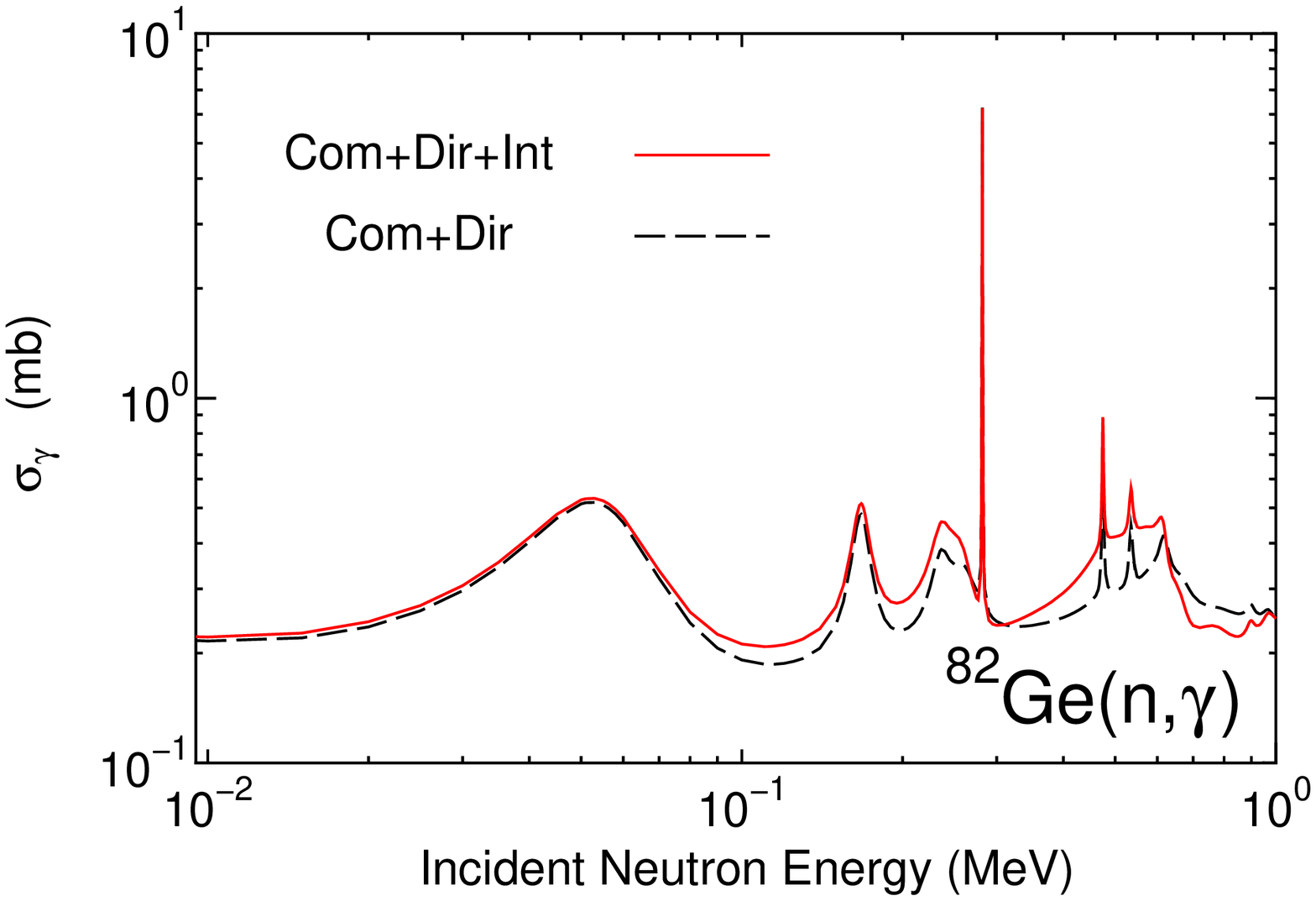}
\caption{Neutron capture cross sections of $^{134}$Sn (top), and $^{82}$Ge (bottom) from $E=0.01$ to $1$ MeV. The results of compound process + direct process + interference effect (Com+Dir+Int), compound process + direct process (Com+Dir) are shown by the solid and dashed lines, respectively. The resonance parameters of $^{134}$Sn and $^{82}$Ge are taken from TENDL~\cite{TENDL}.}
\label{capture}
\end{figure}

Figure \ref{capture} shows the neutron capture cross sections of $^{134}$Sn and $^{82}$Ge. We can see that the interference effect gives a significant change at the first resonance appearing at $E=102.1$ keV for $^{134}$Sn. The changes by the interference effect can be also observed in the other resonances distributing from $E=0.5$ to $1.0$ MeV although the modifications are not as strong as the first one. As seen in $^{13}$C, the interference effect increases the cross section from low energy because the $s$-wave neutron capture into the $3p_{3/2}$ and $3p_{1/2}$ bound states is possible, and the compound and direct processes have a comparable cross section. In case of $^{82}$Ge, because of the multilevel resonances composed of $J=1/2$ and $3/2$ above $E=100$ keV, the interference effect gives complicated changes.

In order to evaluate the effect of the interference effect in an astrophysical condition, we calculate the Maxwellian averaged cross section at $kT=30$ and $300$ keV. The result comparing Com+Dir+Int and Com+Dir at $kT=30$ keV are shown in Table~\ref{mac30}. By introducing the interference effect, the cross sections are increased for nuclei studied in this work. In particular, the neutron capture cross sections for $^{134}$Sn changes considerably as compared to the other two nuclei. The Maxwellian averaged cross section at $kT=300$ keV is shown in Table \ref{mac300}. Contrary to the results obtained for $kT=30$ keV, the interference effect decreases the neutron capture cross sections of  $^{13}$C and $^{134}$Sn. On the other hand, the variation of $^{82}$Ge by including the interference effect is not changed significantly from that for $kT=30$ keV.

The difference of the interference effect at $kT=30$ and $300$ keV can be understood by the followings; Because the Maxwell-Boltzmann distribution has an arched curve with the peak at $E=kT$, the cross section is increased by the interference effect if the temperature is sufficiently smaller than the first resonance. However, if the Maxwell-Boltzmann distribution shifts to higher energy as the temperature increases, the cross section is affected not only by the constructive part but also by the destructive part of the interference effect. This case is realized in $^{13}$C and $^{134}$Sn. In case of $^{82}$Ge, because there are a lot of resonances from $E=0.01$ to $1$ MeV, the interpretation of the difference between $kT=30$ and $300$ keV is not simple. However, the reduction of the capture cross section above $E=700$ keV results in the smaller Maxwellian averaged cross section at $kT=300$ keV than that at $kT=30$ keV.

\begin{table}
\begin{tabular}{c|ccc}
\hline\hline
                        & $^{13}$C & $^{82}$Ge & $^{134}$Sn \\
\hline
Com+Dir+Int & 0.0700 & 0.3574 & 0.3511\\
Com+Dir       & 0.0692 & 0.3412 & 0.2690\\
\hline
Variation (\%) & +1.1 & +4.7 & +30 \\
\hline\hline
\end{tabular}
\caption{The Maxwellian averaged cross sections at $kT=30$ keV of $^{13}$C, $^{82}$Ge, and $^{134}$Sn in unit of mb. The results of compound process + direct process + interference effect (Com+Dir+Int) and compound process + direct process (Com+Dir) are listed. The variation is calculated by $(\sigma(\rm{Com+Dir+Int})/\sigma(\rm{Com+Dir})-1)\times 100$.}
\label{mac30}
\end{table}
\begin{table}
\begin{tabular}{c|ccc}
\hline\hline
                        & $^{13}$C & $^{82}$Ge & $^{134}$Sn \\
\hline
Com+Dir+Int & 0.0776 & 0.6367 & 0.2566\\
Com+Dir        & 0.0780 & 0.6115 & 0.2591\\
\hline
Variation (\%) & -0.5 & +4.1 & -1.0 \\
\hline\hline
\end{tabular}
\caption{Same as Table \ref{mac30} but at $kT=300$ keV.}
\label{mac300}
\end{table}

\section{Summary}
\label{sec-4}
We have estimated the interference effect of the neutron capture cross sections on the neutron-rich nuclei $^{82}$Ge and $^{134}$Sn, and light nucleus $^{13}$C. The neutron capture cross section of the compound process was calculated by the resonance parameters and that of the direct process was calculated by the potential model. The phase problem in the interference effect was chosen so that the interference effect became constructive below the resonance and destructive above it.

In case of $^{13}$C, the neutron capture cross section from $E=10^{-8}$ to $1$ MeV. It was found that the interference effect increased the neutron capture cross section significantly. However, its effect was relatively small around the first resonance. In case of $^{134}$Sn and $^{82}$Ge, the interference  effect was clearly observed around the resonances.

The Maxwellian averaged cross sections at $kT=30$ and $300$ keV were calculated. The changes of the neutron capture cross section by the interference effect were sensitive in case of $kT=30$ keV for nuclei studied in this work. When the temperature was set to $kT=300$ keV, the interference effect became weak because of the constructive and destructive role of the interference effect of neighboring resonances.

This work suggests that the interference effect may have a large impact on neutron-rich nuclei as well as other light nuclei. However, in order to put forward the interference effect in the neutron capture reactions practically, we must solve the problems about the use of artificial resonance parameters, the phase problem, and the partial $\gamma$-decay width. Thus, new experimental measuring cross section and $\gamma$-ray energy and its strength around resonances is required and it will provide an further understanding of the interference effect.

\section*{Acknowledgment}
This work is supported by YUGOKENKYU project at Japan Atomic Energy Agency.

\end{document}